\documentclass[12pt]{article}
\usepackage{epsf}
\title{ {\bf The radiative $h^0 (A^0)\rightarrow \gamma\gamma$  decays
in the two Higgs doublet model. }}

\author{\vspace{1cm}\\
        {\bf E. O. Iltan}
        \thanks{E-mail address:
        eiltan@heraklit.physics.metu.edu.tr}
 \\
        Physics Department, Middle East Technical University \\
        Ankara, Turkey\\}
\date{}
\begin{document}
\setlength{\baselineskip}{24pt}
\maketitle
\setlength{\baselineskip}{7mm}
\begin{abstract}
We study the radiative $h^0 (A^0)\rightarrow \gamma\gamma$ decays
in the framework of the general two Higgs doublet model, the
so-called model III. We estimate the decay width of the processes
at the order of magnitude of $10^{-2}\, GeV$ for the intermediate
values of the Yukawa couplings. We observe that the analysis of
the two photon system provide a comprehensive information for the
complex nature of the Yukawa couplings and therefore, the source
of the CP violation.
\end{abstract}
\thispagestyle{empty}
\newpage
\setcounter{page}{1}
%%%
%%%
\section{Introduction}
The radiative decays $S\rightarrow \gamma \gamma$ (S: scalar,
pseudo-scalar particle) are interesting to analyze since the
output is the two photon system and it is rich due to two
different polarizations of photons, namely the parallel and the
perpendicular ones. This leads to two different amplitudes, CP odd
and CP even, and the measurement of the photon spin polarization
provides comprehensive information about the model parameters
existing in the amplitudes. Besides this, such decays are the
strong candidates to study the direct or indirect CP violation.

There are various studies on the two photon system in the
literature. The decays of mesons into two photons have been
analyzed theoretically \cite{Theo} in various models and
experimentally \cite{Expr} in different collaborations. The decay
of the standard model (SM) Higgs boson $H^0$ into two photons is
another candidate for the two photon system as an output and it
can provide a valuable information to understand the electroweak
symmetry breaking and to predict the model parameters
\cite{H0gamgam1,H0gamgam2}. In the SM, the $H^0\rightarrow
\gamma\gamma$ decay exists at the loop level with the help of all
massive particles, which have non-zero electromagnetic charge,
running in the loop. The most important contributions come from
the top quark and W boson. The branching ratio (BR) has been
predicted at the order of the magnitude of $10^{-3}$ in the
framework of SM. The addition of the dimension six and dimension
eight fermionic and bosonic operators enhances the SM width of
this decay up to one order of magnitude \cite{H0gamgam2}.

In the two Higgs doublet model (2HDM) which is the minimal
extension of the SM, there appears the well known new Higgs
scalars $h^0$ and $A^0$ in addition to the SM one, $H^0$. The two
photon decays of these new particles can provide a considerable
information about the model parameters since they occur at least
at the one loop level where all the charged fermions run in the
loop. On the other hand it is possible to study the CP violating
effects since the output is the two photon system and it has two
different states, CP odd and CP even.

In the present work we analyze the two photon decays of new Higgs
bosons $h^0$ and $A^0$, $h^0 (A^0)\rightarrow \gamma\gamma$, in
the framework of the model III version of the 2HDM. Since the
charged fermions, which are running in the loop, are the source of
the processes under consideration there exist various Yukawa
couplings which are diagonal in flavor indices and complex in
general. In our analysis we take into account only the Yukawa
couplings corresponding to the heaviest fermions in each group of
particles, up-quarks, down-quarks, charged leptons. We also use
the restrictions obtained in the literature for the existing
Yukawa couplings and we estimate the decay width of the processes
at the order of magnitude of $10^{-2}\, GeV$ for the intermediate
values of the Yukawa couplings. On the other hand, these couplings
are complex in general and a new parameter carrying the imaginary
part of the coupling arises. However, the analysis of the CP even
$A^+_{h^0 (A^0)}$ and CP odd $A^-_{h^0 (A^0)}$ amplitudes, which
correspond to the parallel and perpendicular spin polarization of
photon system, shows that the strength of the complexity of the
couplings can be predicted by measuring the relative magnitudes of
CP odd and CP even amplitudes. In addition to this the measurement
these amplitudes and the $BR$ of the processes provide wide
information about the masses $m_{h^0}$ and $m_{A^0}$.

The paper is organized as follows: In Section II, we present the
theoretical expression for the decay widths of  $h^0\rightarrow
\gamma\gamma$ and $A^0 \rightarrow \gamma\gamma$ decays, in the
framework of the model III. Section 3 is devoted to discussion and
our conclusions.
%%%
%%%
\section{The radiative $h^0 (A^0)\rightarrow \gamma\gamma$  decays in
the two Higgs doublet model}
The Higgs bosons  $h^0$ and $A^0$ can decay into two photon system
with the help of the fermion loop, which includes all possible
quarks and down leptons. The interaction between the scalar bosons
and the fermions are derived by the Yukawa lagrangian and the most
general one in the 2HDM reads
\begin{eqnarray}
{\cal{L}}_{Y}&=&\eta^{U}_{ij} \bar{Q}_{i L} \tilde{\phi_{1}} U_{j
R}+ \eta^{D}_{ij} \bar{Q}_{i L} \phi_{1} D_{j R}+
\xi^{U\,\dagger}_{ij} \bar{Q}_{i L} \tilde{\phi_{2}} U_{j R}+
\xi^{D}_{ij} \bar{Q}_{i L} \phi_{2} D_{j R} \nonumber \\ &+&
\eta^{E}_{ij} \bar{l}_{i L} \phi_{1} E_{j R}+ \xi^{E}_{ij}
\bar{l}_{i L} \phi_{2} E_{j R} + h.c. \,\,\, , \label{lagrangian}
\end{eqnarray}
where $L$ and $R$ denote chiral projections $L(R)=1/2(1\mp
\gamma_5)$, $\phi_{i}$ for $i=1,2$, are the two scalar doublets,
$\bar{Q}_{i L}$ are left handed quark doublets, $U_{j R} (D_{j
R})$ are  right handed up (down) quark singlets, $l_{i L}$ ($E_{j
R}$) are lepton doublets (singlets), with family indices $i,j$.
Here the Yukawa matrices $\xi^{U,D}_{ij}$ and $\xi^{E}_{ij}$ have
in general complex entries. The possible mixing between neutral
Higgs bosons $H^0$ and $h^0$ is switched off by choosing the
parametrization for $\phi_{1}$ and $\phi_{2}$ as
\begin{eqnarray}
\phi_{1}=\frac{1}{\sqrt{2}}\left[\left(\begin{array}{c c}
0\\v+H^{0}\end{array}\right)\; + \left(\begin{array}{c c} \sqrt{2}
\chi^{+}\\ i \chi^{0}\end{array}\right) \right]\, ;
\phi_{2}=\frac{1}{\sqrt{2}}\left(\begin{array}{c c} \sqrt{2}
H^{+}\\ H_1+i H_2 \end{array}\right) \,\, , \label{choice}
\end{eqnarray}
where only $\phi_{1}$ has a vacuum expectation value;
\begin{eqnarray}
<\phi_{1}>=\frac{1}{\sqrt{2}}\left(\begin{array}{c c}
0\\v\end{array}\right) \,  \, ; <\phi_{2}>=0 \,\, ,
\label{choice2}
\end{eqnarray}
and by considering the gauge and $CP$ invariant Higgs potential
which spontaneously breaks  $SU(2)\times U(1)$ down to $U(1)$  as
\begin{eqnarray}
V(\phi_1, \phi_2 )&=&c_1 (\phi_1^+ \phi_1-v^2/2)^2+ c_2 (\phi_2^+
\phi_2)^2 \nonumber \\ &+&  c_3 [(\phi_1^+ \phi_1-v^2/2)+ \phi_2^+
\phi_2]^2 + c_4 [(\phi_1^+ \phi_1)
(\phi_2^+ \phi_2)-(\phi_1^+ \phi_2)(\phi_2^+ \phi_1)] \nonumber \\
&+& c_5 [Re(\phi_1^+ \phi_2)]^2 + c_{6} [Im(\phi_1^+ \phi_2)]^2
+c_{7} \, , \label{potential}
\end{eqnarray}
with constants $c_i, \, i=1,...,7$. This is the case where the SM
(new) particles are collected in the first (second) doublet.
Notice that, in the following we will replace $\xi^{U,D,E}_{ij}$
by $\xi^{U,D,E}_{N,ij}$ to emphasize that the couplings are
related to the neutral interactions.

\begin{figure}[htb]
\vskip 2.0truein \centering \epsfxsize=3.2in
\leavevmode\epsffile{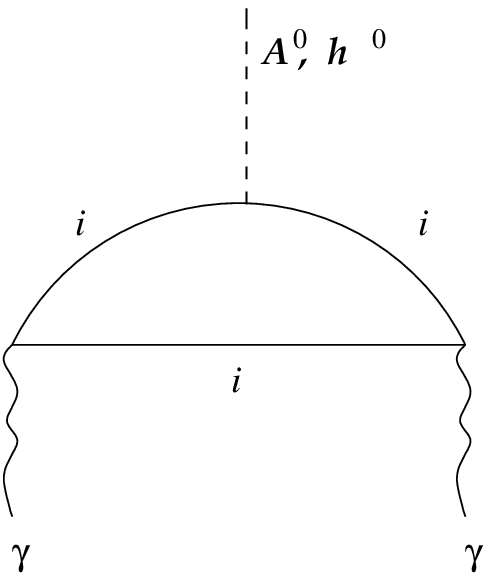} \vskip 1.0truein \caption[]{One loop
level diagrams contribute to $\Gamma (h^0 (A^0))\rightarrow
\gamma\gamma$ decay  in the model III version of 2HDM. Solid lines
represent the charged fermions denoted by $i$, dashed lines
represent the $h^0$ and $A^0$ fields and curly lines represent the
photon fields.} \label{fig1}
\end{figure}

Now, we consider the radiative decays of the Higgs bosons $h^0$
and $A^0$, $h^0(A^0) \rightarrow \gamma\gamma$. These processes
occur with the internal fermion flow Fig. (\ref{fig1}) and the
decay amplitudes strongly depend on the Yukawa couplings of the
Higgs-fermion interactions. The decay amplitude for these
radiative decays can be written in terms of two Lorentz
structures:
\begin{eqnarray}
M(S\rightarrow \gamma\gamma)=A_S^+ \,F_{\mu\nu}\,F^{\mu\nu}+i\,
A_S^- \,F_{\mu\nu}\,\tilde{F}^{\mu\nu}, \label{Ampl1}
\end{eqnarray}
with
$\tilde{F}^{\mu\nu}=\frac{1}{2}\,\epsilon_{\mu\nu\alpha\beta}\,
F^{\alpha\beta}$ and $A_S^+$ ($A_S^-$) are CP even (odd)
amplitude:
\begin{eqnarray}
A^+_S &=& \frac{c_1^{S}}{4\,\pi\,m^2_{S}}\,\sum _i\,m_i\, N_i\,
Q_i^2\,g^{S\,+}_{ii}\,\{-2+(-1+4\,x_i^{S})\,(Li_2[f_+]+Li_2[f_-])\}
\, , \nonumber \\
A^-_S &=& \frac{c_1^{S}}{4\,\pi\,m^2_{S}}\,\sum _i\,m_i\, N_i\,
Q_i^2 \, g^{S\, -}_{ii} (Li_2[f_+]+Li_2[f_-]) , \label{Ampl12}
\end{eqnarray}
where $S$ is the scalar $h^0$ or $A^0$, $c_1^{h^0\, (A^0)}=-(I)
\sqrt{2}\, \alpha_e$, $\alpha_e=\frac{1}{137}$,  $g^{h^0\,
\pm}_{ii}=(\xi^{F}_{N,ii} \pm \xi^{F\,*}_{N,ii})$,
$g^{A^0\,\pm}_{ii}=g^{h^0\,\mp}_{ii}$, $F= U\,(D;E)$ for
$i=u,c,t\,\, (d,s,b;\, e,\mu,\tau)$ , $N_i$ is the color factor,
$N_i=3\, (1)$ for $i=u,c,t,d,s,b \,\, (e,\mu,\tau)$, and $Q_i$ is
the charge $Q_i= \frac{2}{3}\,(\frac{-1}{3};-1)$ for $i=u,c,t\,\,
(d,s,b; \, e,\mu,\tau)$. Here the functions $Li_2[z]$ are the
PolyLogarithmic function $Li_2[z]=\int_{z}^0\,
\frac{Log[1-t]}{t}\, dt$ and $f_\pm=\frac{2}{1\pm\sqrt{1-4\,x_i}}$
with $x_i=\frac{m_i^2}{m^2_{S}}$. Furthermore, we parametrize the
complex Yukawa coupling $\xi^{U(D,E)}_{N,ii}$ as
\begin{equation}
\xi^{U,D,E}_{N,i i}=|\xi^{U,D,E}_{N,ii}|\, e^{i \theta_{ii}} \, .
\label{xicomplex}
\end{equation}
Finally the decay width of the processes can be obtained by using
the expression
\begin{eqnarray}
\Gamma_S=\frac{m_S^3}{16\,\pi}\, (|A^+_S|^2+|A^-_S|^2)
\label{Dwdt}
\end{eqnarray}

In the decays $h^0\rightarrow \gamma\gamma$ and $A^0\rightarrow
\gamma\gamma$, the outgoing photons are in one of the possible
states given by $F^{\mu\nu}\,F_{\mu\nu}$ and
$F^{\mu\nu}\,\tilde{F}_{\mu\nu}$ which are even and odd in CP,
respectively. These states corresponds to the parallel
($\epsilon_1 . \epsilon_2$) and perpendicular ($\epsilon_1\times
\epsilon_2$) spin polarizations of photons. Eq. (\ref{Ampl12})
shows that the decay amplitude is CP even (odd) for the
$h^0\rightarrow \gamma\gamma$ ($A^0\rightarrow \gamma\gamma$)
decay in the case of real couplings, when the one loop
calculations are taken into account. The strength of the complex
part in the Yukawa couplings can be estimated by detecting the CP
odd (even) amplitude relative to CP even (odd) one, in other words
by studying  the ratio $\frac{|A^+_S|^2}{|A^-_S|^2}$, for the
$h^0\rightarrow \gamma\gamma$ ($A^0\rightarrow \gamma\gamma$)
decay. Therefore the precise determination of the photon
polarization in the experiments give considerable information
about the free parameters of the theory under consideration.
%%%
%%%
\section{Discussion}
This section is devoted to the analysis of the decay width and the
amplitudes for the  CP even and CP odd photon states of the
processes $h^0 \rightarrow \gamma \gamma$ and $A^0 \rightarrow
\gamma \gamma$. These processes occur at least at the loop order
with the help of the internal charged fermions Fig. \ref{fig1}.
They are driven by the  diagonal Yukawa couplings
$\xi^{U}_{N,ii}$, ($i$=up quarks), $\xi^{D}_{N,ii}$, ($i$=down
quarks) and $\xi^{E}_{N,ii}$, ($i$=down leptons), which are the
free parameters of the model used, and they should be restricted
by using the appropriate experimental measurements.  At first, we
assume that the Yukawa couplings $\xi^{U}_{N,ii},\, i=u,c$,
$\xi^{D}_{N,ii},\, i=d,s$ and  $\xi^{E}_{N,ii},\, i=e,\mu$, are
small compared to $\xi^{U}_{N,tt}$, $\xi^{D}_{N,bb}$ and
$\xi^{E}_{N,\tau\tau}$ since the strength of them are related to
the masses of fermions denoted by the indices, similar to the
Cheng-Sher scenerio \cite{Sher}. Second, we use the predictions
for the couplings $\xi^{D}_{N,bb}$, $\xi^{U}_{N,tt}$, obtained by
using the constraint region due to the restriction of the Wilson
coefficient $C_7^{eff}$, which is the effective coefficient of the
operator $O_7 = \frac{e}{16 \pi^2} \bar{s}_{\alpha} \sigma_{\mu
\nu} (m_b R + m_s L) b_{\alpha} {\cal{F}}^{\mu \nu}$ (see
\cite{Alil1} and references therein), in the region $0.257 \leq
|C_7^{eff}| \leq 0.439$. Here upper and lower limits were
calculated using the CLEO measurement \cite{cleo2}
\begin{eqnarray}
BR (B\rightarrow X_s\gamma)= (3.15\pm 0.35\pm 0.32)\, 10^{-4} \,\,
, \label{br2}
\end{eqnarray}
and all possible uncertainties in the calculation of $C_7^{eff}$
\cite{Alil1}. The above restriction ensures to get upper and lower
limits for $\xi^{D}_{N,bb}$, $\xi^{U}_{N,tt}$ (see \cite{Alil1}
for details). At this stage we will make the parametrization
\begin{equation}
\xi^{U,D,E}_{N,ij}= \sqrt{\frac{4 G_F}{\sqrt {2}}}
\bar{\xi}^{U,D,E}_{N,ij} \, , \label{ksipar}
\end{equation}
where $G_F$ is the fermi constant and use it in the following.

Our numerical calculations are based on the choice intermediate
values of the couplings for $C_7^{eff}>0$,
$\bar{\xi}^{D}_{N,bb}=30\,m_b$, $\bar{\xi}^{U}_{N,tt}\sim 0.0025\,
GeV$, respecting the constraints mentioned above. The coupling
$\bar{\xi}^{E}_{N,\tau \tau}$ is the other one which should be
restricted and we use the prediction obtained in the analysis of
the decay $\tau\rightarrow \mu \, \bar{\nu_i}\, \nu_i$,
$i=e,\mu,\tau$, in \cite{taumununu}. In this work the $BR$ is
estimated at the order of the magnitude of $10^{-6}-10^{-4}$ for
the range of the couplings, $\bar{\xi}^{E}_{N,\tau \tau}\sim
30-100 \, GeV$ and $\bar{\xi}^{E}_{N,\tau \mu}\sim 10-25\, GeV$.
In the present work, for the coupling $\bar{\xi}^{E}_{N,\tau
\tau}$,  we take the optimum value $\bar{\xi}^{E}_{N,\tau
\tau}\sim 30\, GeV$. In addition to this we consider the general
case that the couplings are complex (see eq. \ref{xicomplex})

\begin{figure}[htb]
\vskip -3.0truein \centering \epsfxsize=6.8in
\leavevmode\epsffile{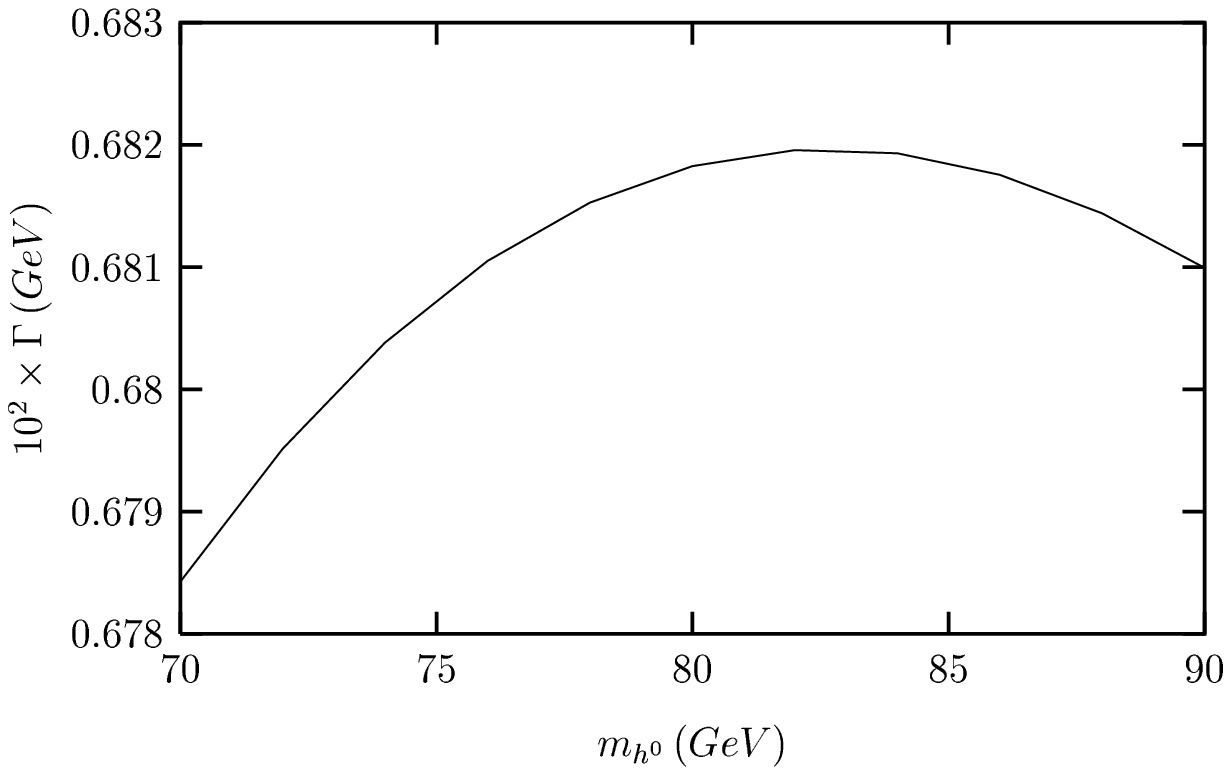} \vskip -3.0truein
\caption[]{The $m_{h^0}$ dependence of the $\Gamma$ for the decay
$h^0 \rightarrow \gamma \gamma$, for the intermediate value of the
complexity parameter $sin\theta_{ii}=0.5$ (i=t-quark, b-quark,
$\tau$-lepton). } \label{BRh0gamgamh00}
\end{figure}

\begin{figure}[htb]
\vskip -3.0truein \centering \epsfxsize=6.8in
\leavevmode\epsffile{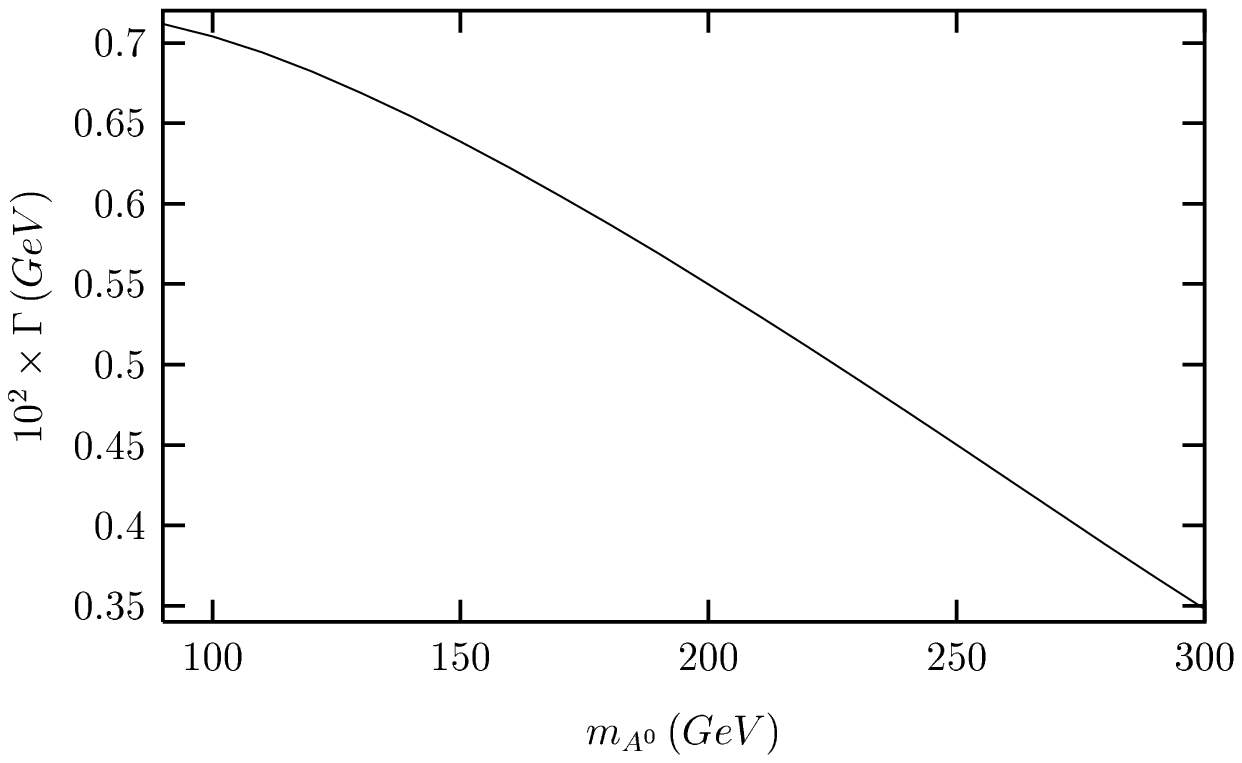} \vskip -3.0truein
\caption[]{The $m_{A^0}$ dependence of the $\Gamma$ for the decay
$A^0 \rightarrow \gamma \gamma$, for the intermediate value of the
complexity parameter $sin\theta_{ii}=0.5$ (i=t-quark, b-quark,
$\tau$-lepton).  } \label{BRA0gamgamA0}
\end{figure}

In Fig. \ref{BRh0gamgamh00} (\ref{BRA0gamgamA0}), we present
$m_{h^0}\, (m_{A^0})$ dependence of the decay width $\Gamma$ of
the $h^0 (A^0) \rightarrow \gamma \gamma$ decay, for the
intermediate value of the complexity parameter
$sin\theta_{ii}=0.5$ ($i=t$-quark, $b$-quark, $\tau$-lepton) of
three Yukawa couplings. The Figure \ref{BRh0gamgamh00}
(\ref{BRA0gamgamA0}) shows that the $\Gamma$ is at the order of
the magnitude of $10^{-3}-10^{-2}\,GeV$ for the decay $h^0
\rightarrow \gamma \gamma$ ($A^0 \rightarrow \gamma \gamma$) and
it is not so much sensitive to the mass of $h^0$ in the given mass
interval (decreases with the increasing values of the mass of
$A^0$). With the possible measurement of this radiative decay a
considerable information can be obtained for the mass of the
scalar bosons, especially for $m_{A^0}$.

\begin{figure}[htb]
\vskip -3.0truein \centering \epsfxsize=6.8in
\leavevmode\epsffile{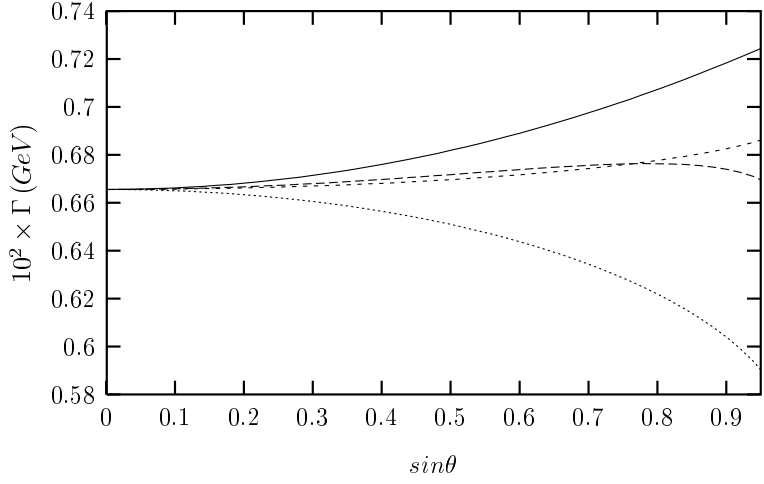} \vskip -3.0truein
\caption[]{The $sin\theta$ dependence of the $\Gamma$ for the
decay $h^0 \rightarrow \gamma \gamma$ and $m_{h^0}=85\, GeV$. The
solid line represents the case where all the Yukawa couplings are
complex and $sin\theta_{bb}=sin\theta_{tt}=sin\theta_{\tau\tau}$,
the dashed (small dashed; dotted) line represents the case where
$sin\theta_{tt}=sin\theta_{\tau\tau}=0$
($sin\theta_{bb}=sin\theta_{\tau\tau}=0$;
$sin\theta_{bb}=sin\theta_{tt}=0$).} \label{BRh0gamgamsin}
\end{figure}

\begin{figure}[htb]
\vskip -3.0truein \centering \epsfxsize=6.8in
\leavevmode\epsffile{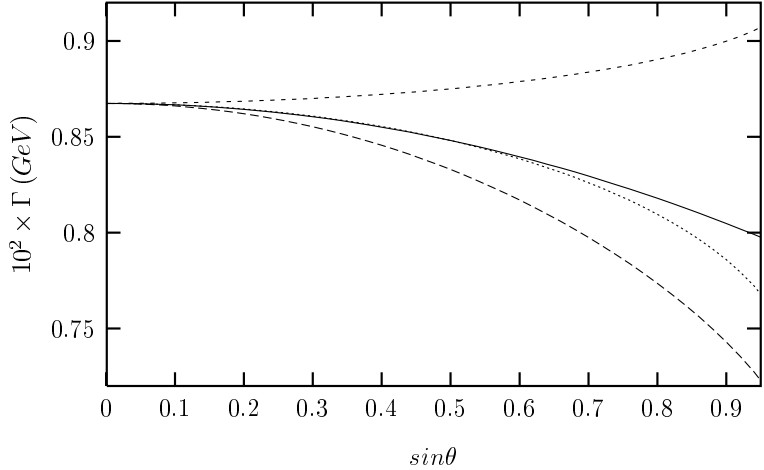} \vskip -3.0truein
\caption[]{The same as Fig. \ref{BRh0gamgamsin} but for the decay
$A^0 \rightarrow \gamma \gamma$ and $m_{A^0}=90\, GeV$.}
\label{BRA0gamgamsin}
\end{figure}

Fig. \ref{BRh0gamgamsin} (\ref{BRA0gamgamsin}) is devoted to the
$sin\theta$ dependence of the $\Gamma (h^0 (A^0) \rightarrow
\gamma \gamma)$ for $m_{h^0}=85\, GeV$ ($m_{A^0}=90\, GeV$). The
solid line represents the case where all the Yukawa couplings are
complex and $sin\theta_{bb}=sin\theta_{tt}=sin\theta_{\tau\tau}$,
the dashed (small dashed; dotted) line represents the case where
$sin\theta_{tt}=sin\theta_{\tau\tau}=0$
($sin\theta_{bb}=sin\theta_{\tau\tau}=0$;
$sin\theta_{bb}=sin\theta_{tt}=0$). We observe that the increasing
values of $sin\theta_{bb}$ ($sin\theta_{tt}$,
$sin\theta_{\tau\tau}$) causes to enhance (weakly supress,
supress) the $\Gamma$ in the $h^0 \rightarrow \gamma \gamma$
decay.  For the $A^0 \rightarrow \gamma \gamma$ decay, the
increasing values of $sin\theta_{bb}$ ($sin\theta_{tt}$,
$sin\theta_{\tau\tau}$) causes to supress (enhance, supress) the
$\Gamma$. In the case that all the Yukawa couplings are complex
and have the same value, there is an enhancement (suppression) in
the $\Gamma$ for the increasing values of $sin\theta$, for the
$h^0 \rightarrow \gamma \gamma$ ($A^0 \rightarrow \gamma \gamma$)
decay. These results may be effective in the determination of the
complexity of the Yukawa couplings.

\begin{figure}[htb]
\vskip -3.0truein \centering \epsfxsize=6.8in
\leavevmode\epsffile{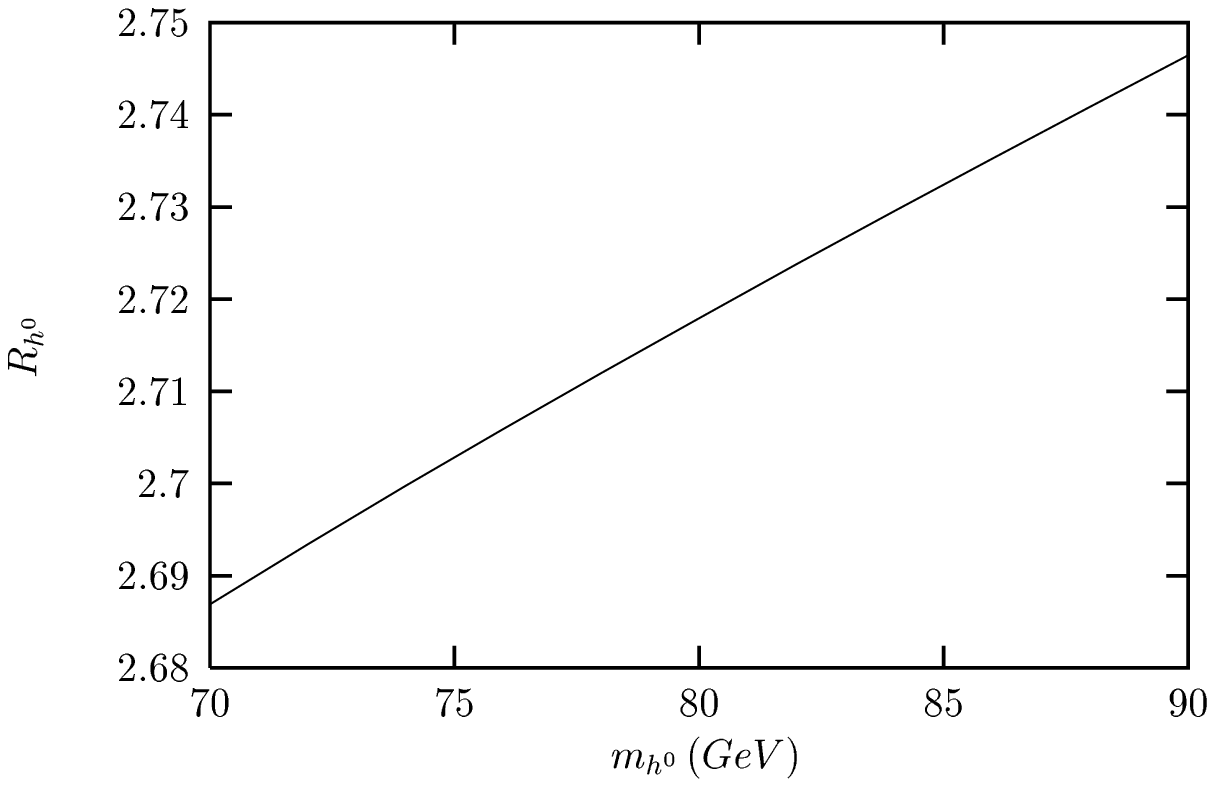} \vskip -3.0truein
\caption[]{The same as Fig. \ref{BRh0gamgamh00} but for the ratio
$R_{h^0}$.} \label{Rh0gamgamh0}
\end{figure}

\begin{figure}[htb]
\vskip -3.0truein \centering \epsfxsize=6.8in
\leavevmode\epsffile{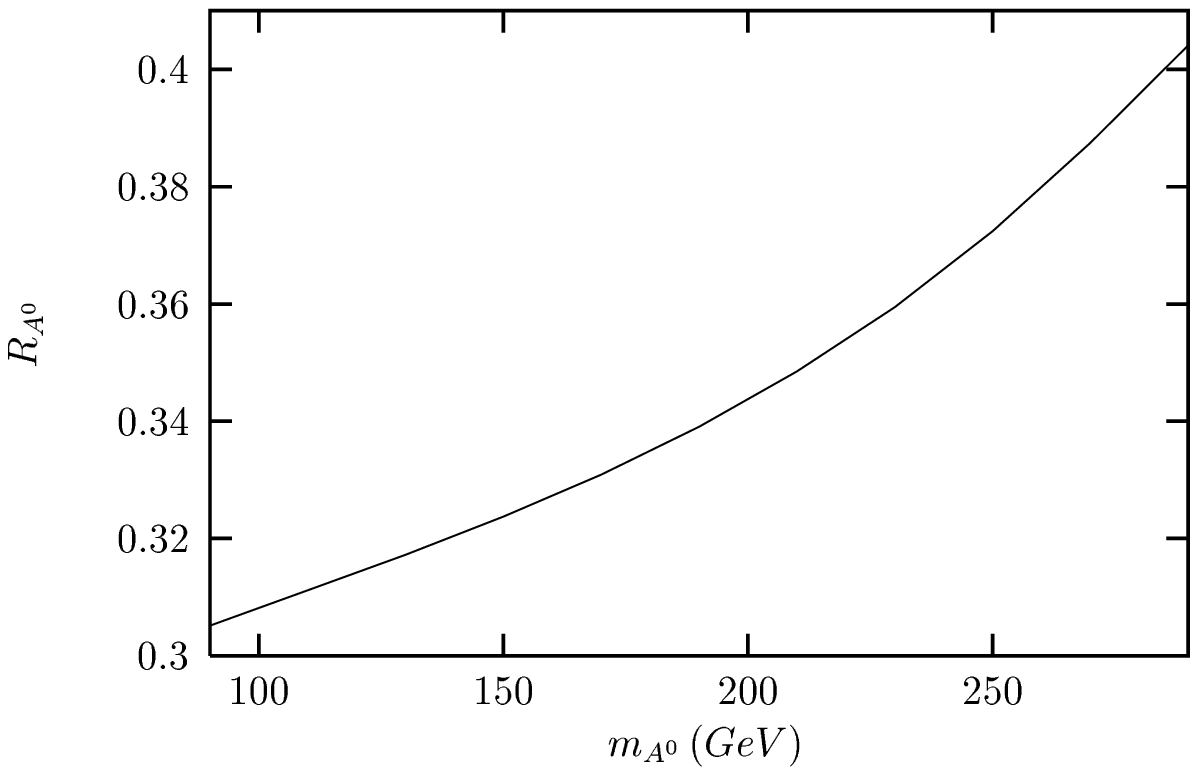} \vskip -3.0truein
\caption[]{The same as Fig. \ref{BRA0gamgamA0} but for the ratio
$R_{A^0}$.} \label{RA0gamgamA0}
\end{figure}

The ratio of CP even and CP odd amplitudes
$R_S=\frac{|A^+_S|^2}{|A^-_S|^2}$ is interesting since it depends
strongly on the complex nature of the Yukawa couplings. In Fig.
\ref{Rh0gamgamh0} (\ref{RA0gamgamA0}), we present $m_{h^0}$
($m_{A^0}$) dependence of the ratio $R_{h^0}$ ($R_{A^0}$) for the
decay $h^0 (A^0) \rightarrow \gamma \gamma$, for the intermediate
value of the parameter $sin\theta_{ii}=0.5$. Figure
\ref{Rh0gamgamh0} (\ref{RA0gamgamA0}) shows that the $R_{h^0}$
($R_{A^0}$) is greater than one (less than one)  for the given
range of mass, for the decay $h^0 \rightarrow \gamma \gamma$ ($A^0
\rightarrow \gamma \gamma$) and increases with the increasing
values of the mass $h^0$ ($A^0$).

\begin{figure}[htb]
\vskip -3.0truein \centering \epsfxsize=6.8in
\leavevmode\epsffile{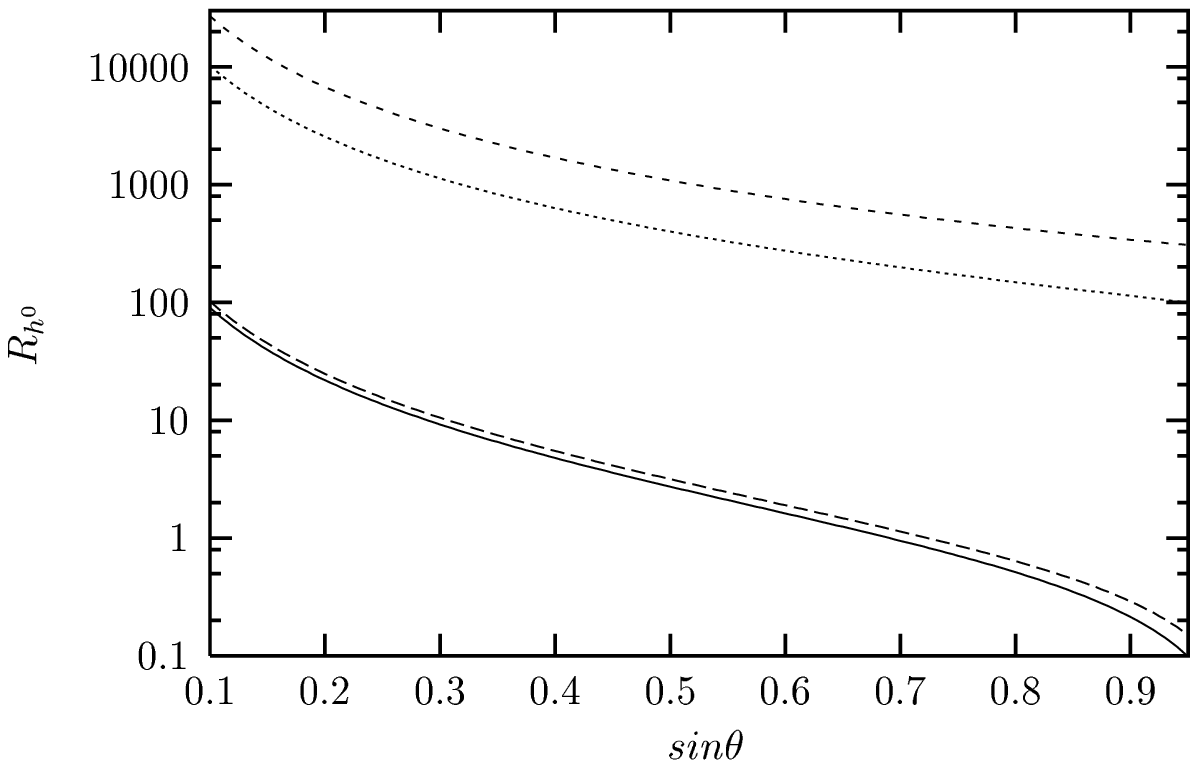} \vskip -3.0truein
\caption[]{The same as Fig. \ref{BRh0gamgamsin} but for the ratio
$R_{h^0}$.} \label{Rh0gamgamsin}
\end{figure}

\begin{figure}[htb]
\vskip -3.0truein \centering \epsfxsize=6.8in
\leavevmode\epsffile{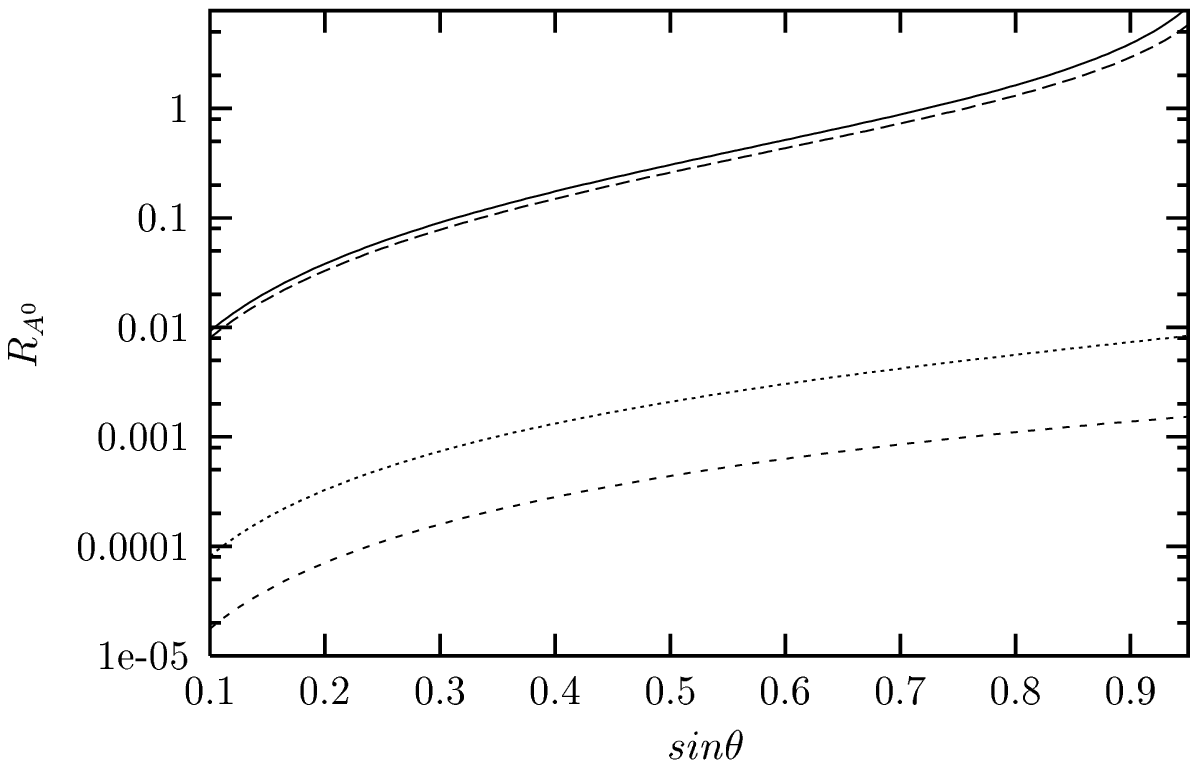} \vskip -3.0truein
\caption[]{The same as Fig. \ref{BRA0gamgamsin} but for the ratio
$R_{A^0}$.} \label{RA0gamgamsin}
\end{figure}

Fig.\ref{Rh0gamgamsin} (\ref{RA0gamgamsin}) represents  the
$sin\theta$ dependence of the ratio $R_{h^0}$ ($R_{A^0}$) for
$m_{h^0}=85\, GeV$ ($m_{A^0}=90\, GeV$). The solid line represents
the case where all the Yukawa couplings are complex and
$sin\theta_{bb}=sin\theta_{tt}=sin\theta_{\tau\tau}$, the dashed
(small dashed; dotted) line represents the case where
$sin\theta_{tt}=sin\theta_{\tau\tau}=0$
($sin\theta_{bb}=sin\theta_{\tau\tau}=0$;
$sin\theta_{bb}=sin\theta_{tt}=0$). We observe that the ratio
$R_{h^0}$ ($R_{A^0}$) is strongly sensitive to the complexity of
the Yukawa couplings. For the real couplings $|A^-_S|^2$ vanishes
and the ratio $R_{h^0}$ tends to infinity. In such a case the only
possible state of two photon system is the parallel polarization.
The measurement of the perpendicular polarization for the photon
system is a sign for the complexity of the Yukawa couplings. The
ratio $R_{h^0}$ decreases with the increasing values of
$sin\theta$ for all Yukawa couplings. This ratio is larger for the
case that only the Yukawa coupling $\xi^{U}_{N,tt}$ (or
$\bar{\xi}^{E}_{N,\tau \tau}$) is complex. This observation can
make it possible to determine the relative strengths of the
complexity of different Yukawa couplings. On the other hand, the
ratio $R_{A^0}$ vanishes for the real coupling and only the
perpendicular polarization state exists for the two photon system.
Therefore, for the decay $A^0 \rightarrow \gamma \gamma$, the
measurement of the parallel polarization for the photon system is
a sign for the complexity of the Yukawa couplings. The ratio
$R_{A^0}$ increases with the increasing values of $sin\theta$ for
all Yukawa couplings and it is smaller for the case that only the
Yukawa coupling $\xi^{U}_{N,tt}$ (or $\bar{\xi}^{E}_{N,\tau
\tau}$) is complex. Similar to the $h^0 \rightarrow \gamma \gamma$
decay, this is informative in the determination of  the relative
strengths of the complexity of different Yukawa couplings.

Now we would like to summarize our results:

\begin{itemize}
\item  We predict the decay width $\Gamma$ of the decay $h^0 (A^0)
\rightarrow \gamma \gamma$ at the order of the magnitude of
$10^{-3}-10^{-2}\,GeV$ and it is not so much sensitive to the mass
of $h^0$ in the given range (decreases with the increasing values
of the mass of $A^0$) Therefore  its measurement can provide
valuable information about the $h^0 \, (A^0)$ masses and the
Yukawa couplings.

\item  We study the ratio $R_{h^0 (A^0)}$ which is constructed
using the  CP even  $A^+_{h^0 (A^0)}$ and CP odd  $A^-_{h^0
(A^0)}$ amplitudes. For the case of real couplings $R_{h^0}$
($R_{A^0}$) tends to infinity (zero) and the only possible state
of two photon system is the parallel (perpendicular) polarization.
The amount of complexity of the Yukawa couplings can explain the
relative magnitude of the perpendicular (parallel) polarization.
Therefore the measurement of the photon polarization of the decay
$h^0 (A^0) \rightarrow \gamma \gamma$ gives a considerable
information about the complex nature of the Yukawa couplings and
strength of their relative complexity.
\end{itemize}

Therefore, the experimental analysis of these radiative decays
would ensure strong information about the scalar boson masses and
the Yukawa couplings existing in the new physics beyond the SM.

\section{Acknowledgement}
This work has been supported by the Turkish Academy of Sciences in
the framework of the Young Scientist Award Program.
(EOI-TUBA-GEBIP/2001-1-8)

\newpage
\newpage
\end{document}